\documentclass{revtex4}

\usepackage{amsmath}    
\usepackage{graphicx}   
\usepackage{verbatim}   
\usepackage{color}      
\usepackage{subfigure}  
\usepackage{hyperref}   
\usepackage{amssymb}
\usepackage{latexsym}
\usepackage{array}
\usepackage[activeacute,english]{babel}
\usepackage[latin1]{inputenc}
\usepackage{lineno}
\usepackage{latexsym}
\def\be{\begin{eqnarray}}
\def\ee{\end{eqnarray}}
\def\nn{\nonumber}
\newcommand{\bq}{\begin{equation}}
\newcommand{\eq}{\end{equation}}
\newcommand{\bc}{\begin{center}}
\newcommand{\ec}{\end{center}}

\def\n{\mbox{\bf n}}
\def\D{\mathcal{D}}

\begin{document}

\title{Induced magnetic moment for a spinless charged particle in the thin-layer approach}

\author{F. T. Brandt}
\email{fbrandt@usp.br}
\affiliation{Instituto de F\'\i sica, Universidade de S\~{a}o Paulo, S\~{a}o Paulo, Brazil}                

\author{ J. A. S\'anchez-Monroy}
\email{antosan@gmail.com}
\affiliation{Instituto de F\'\i sica, Universidade de S\~{a}o Paulo, S\~{a}o Paulo, Brazil}                

\begin{abstract}
We determine the effective dynamics for a spinless charged particle, in the presence of electromagnetic
fields, constrained to a space curve. We employ the thin-layer procedure and a perturbative expansion for
the Schrödinger equation is derived. We find that the first-order term in the perturbative
expansion couples the dynamics of the normal and tangent degrees of freedom. However, there is 
always a gauge transformation that allows decouple the dynamics. We find that the effective Schrödinger
equation in the curve contains an induced Zeeman coupling, independent of the curvature and torsion,
which has not been previously reported. This coupling is characteristic of reducing the dimension by 
two and allows to identify an induced magnetic moment.
\end{abstract}

\maketitle

\section{Introduction}

When the motion of a particle is confined to a low-dimensional space their quantum behavior is strongly affected.
As was found earlier, the particle experiences a quantum potential that is a function of the intrinsic and extrinsic curvatures of
the low-dimensional space in which the particle is confined \cite{Jesen1971,Costa1981}.
The study of the effects of the curvature-induced quantum potential has been the subject
of intensive research \cite{Jaffe2003,Atanasov2007a,Atanasov2008a,Atanasov2009,Atanasov2012a,FerrarriG2008,OrtixCarmine2010,Chaplik2004,
Ono2009,Ono2010,Onoe2009a}. From the theoretical perspective, it was studied the electronic
properties and bound-state formation in curved nanostructures \cite{OrtixCarmine2010}, geometry-induced
charge separation on helicoidal ribbon \cite{Atanasov2009}, curvature-induced p-n junctions in bilayer
graphene \cite{Joglekar2009}, mechanical-quantum-bit states \cite{Chaplik2004}, effects of periodic
curvature on the electrical resistivity of corrugated semiconductor films \cite{Ono2009,Ono2010} as
well as geometry-driven shift in the Tomonaga-Luttinger liquid \cite{Onoe2009a}. 
{From the experimental point of view, 
Szameit \textit{et al.} found an optical analogue of the geometric potential \cite{Szameit2010} and Onoe \textit{et al.} 
reported in 2012 the first experimental evidence of the geometric potential in a quantum system \cite{Onoe2012}.}
\par
The constraints in the so-called \textit{thin layer quantization} method (also known as \textit{confining potential formalism}) are produced by a potential $V_c$, such that the excitation energies of the particle in a normal direction to the lower dimensional system are much larger than those in the tangential direction, so that one can define an effective dynamics for the constrained system 
{\cite{Jesen1971,Costa1981,Jaffe2003,Maraner1993,Maraner95,Maraner1996,Mitchell2001}}.
\par
{The thin layer quantization was derived in the Ref. \cite{Ferrari2008} 
for the Schrödinger equation in the presence of external electromagnetic fields in a 2D curved surface embedded in a 3D space}. 
It was found that there is no coupling between the surface curvature and the external electromagnetic field potential. 
Using a suitable choice of gauge, the dynamics of the normal and tangent degrees of freedom are separable
\cite{Ferrari2008,Ortix2011}. 
{The Schrödinger equation for a particle in a distorted ring in the presence of a magnetic field embedded in a plane was previously derived in Ref. \cite{Pershin2005a}. }
\par
In addition to the quantum potential, these systems exhibit the emergence of a geometry-induced Yang-Mills
field when the low-dimensional space has a geometric torsion
\cite{Takagi1992,Fujii1993,Maraner1993,Maraner95,Maraner1996,Jaffe2003}. The geometry-induced Yang-Mills
field emerges because the orbital angular momentum of the normal coordinates to the low-dimensional space
couples with the geometric torsion. For an ``inner'' observer (an observer who lives in the low-dimension space),
the orbital angular momentum of the normal coordinates is perceived as an ``intrinsic angular momentum''.
As we shall see the wave function in the low-dimensional space will be a multiplet and its dimension will
depend on any degeneracy that exists in the spectrum of the Hamiltonian that governs the normal degrees of 
freedom \cite{Jaffe2003}. Physical consequences of geometry-induced Yang-Mills fields have been explored in 
the last decade \cite{Maraner1996,Taira2010a,Taira2010b,Taira2010c}. It was studied, as the geometric torsion 
inherent in the quantum ring induced a quantum phase shift \cite{Taira2010a}, persistent current 
flow \cite{Taira2010b} and an Aharonov-Bohm-like conductance oscillation \cite{Taira2010b,Taira2010c}.
\par
It is well known that a particle with intrinsic angular momentum $\mathbf{S}$ (spin), in the presence of a magnetic field, interacts with the magnetic field through a Zeeman coupling
\be\label{InteracBS}
g_s\frac{\mu}{\hbar}\mathbf{B}\cdot \mathbf{S}=\frac{\mathbf{B}\cdot\mathbf{\mathbf{\mu}_s}}{\hbar},
\ee
where the Bohr magneton is $\mu=-\frac{e\hbar}{2m}$, $g_s$ the Landé factor and $\mu_s$ the intrinsic
magnetic moment. The Zeeman coupling naturally appears in the Pauli equation or through the non-relativistic limit of the Dirac equation
\cite{Greinerrelativistic1990}. It is clear that for spinless particles this term should no longer appear. However,
we shall see that when a charged spinless particle is confined to a space curve, in the presence of the
external electromagnetic field, an interaction of the type (\ref{InteracBS}) emerges. It will be possible to
identify that the induced magnetic moment depends on the orbital angular momentum of the normal
coordinates. This interaction reinforces the idea that from the point of view of an ``inner'' observer, the
orbital angular momentum of the normal coordinates is perceived as an ``intrinsic angular momentum''.
\par
The aim of our work is to study the effective dynamics of a non-relativistic spinless
charged particle in the presence of electromagnetic fields constrained on a space curve. This paper is
organized as follows. In Sec. \ref{SectionSCH} we derive the effective dynamics for a non-relativistic
spinless and $1/2$-spin charged particles, in the presence of electromagnetic fields,
constrained on a space curve. Finally, Sec. \ref{Section5KG} contains our conclusions.

\section{Schrödinger equation on a curve in the presence of an electromagnetic field}\label{SectionSCH}
We introduce orthogonal curvilinear coordinates $(s, y^1, y^2)$ in $\mathbb{R}^{3}$, where
a rigidly bounded curve $\mathcal{C}$ is parameterized by $\mathbf{r}(s)$. We assume that $\mathcal{C}$ has a
tubular neighborhood, such that ($i=1,2$)
\begin{equation}
\mathbf{R}(s,y^1,y^2)=\mathbf{r}(s)+y^i\mathbf{n}_i(s),
\end{equation}
where $s$ is the arc-length parametrization of the curve and $\mathbf{n}_i(s)$ denotes
two orthonormal vector to $\mathcal{C}$. The curvatures and the normal fundamental form
are $\alpha_{i} \equiv -{\bf t}\cdot\partial_s\hat{\n}_i$ and $\beta_{ij} \equiv\hat{\n}_i\cdot\partial_s \hat{\n}_j$,
respectively, with $\mathbf{t}=d\mathbf{r}/ds$ a tangent vector to the curve $\mathcal{C}$.
The metric in the curvilinear coordinates is given by $G_{ab}=\partial_{a}\mathbf{R}\cdot
\partial_{b}\mathbf{R}$ with $a,b=\{s,y^1,y^2\}$, so that
\be
G_{ab} =\left(\begin{array}{ll}
\gamma+y^k y^l \beta_{kh} \beta_{l}^{\ h}&
       -y^k \beta_{ki} \\
 -y^k \beta_{kj}  & \delta_{ij}
\end{array}\right)
\ee
where $\gamma=(1-y^{k}\alpha_{k})^2$. The inverse of the metric tensor $G_{ab}$ can be calculated exactly,
\be
G^{ab}= \left(\begin{array}{ll}
\lambda & \lambda y^{k}\beta_{k}^{\ j}\\
\lambda y^{k}\beta_{k}^{\ i} & \delta^{ij}+y^{k}y^{l}\beta_{k}^{\ i}\beta_{l}^{\ j}\lambda\end{array}\right),
\ee
with $\lambda=(\gamma^{-1})$.
We will study a spineless charged particle constrained to a curve. In the absence of an electromagnetic
field, this derivation was accomplished in detail in Refs. \cite{Jaffe2003,Maraner95}. In a manner analogous
to what has been done in Refs. \cite{Ferrari2008,Ortix2011}, for particles constrained to a 2D curved surface
embedded in a 3D space, we employ a  gauge covariant derivative $\nabla_{b}=\partial_{b}-\frac{ie}{\hbar}A_{b}$,
where $A_{b}$ are the components of the vector potential $\mathbf{A}$.
This covariant derivative can be extended to $(3+1)$-dimensions defining a gauge covariant derivative for the
time variable as $\D_{0}= \partial_{t}-ieA_{0}/ \hbar$, with $A_0=-\phi$ and where $\phi$ is the scalar potential.
Thus, the Schr\"{o}dinger equation in the presence of the electromagnetic field is
\be\label{SchElcCov}
i\hbar \D_{0}{\Psi(t,s,y^i)}=H\Psi(t,s,y^i)=\left[-\frac{\hbar^2}{2m|G|^{1/2}}D_a
|G|^{1/2}G^{ab}D_b+V_c(y^i)\right]\Psi(t,s,y^i).
\ee
Following the thin layer procedure we first rescaled the Schrödinger field and the Hamiltonian operator as
\begin{eqnarray}
\chi(t,s,y^i)&\equiv&|G|^{1/4}\Psi(t,s,y^i),\\
\mathcal{H}&\equiv&|G|^{1/4}H|G|^{-1/4}.
\end{eqnarray}
In the thin layer method the constraint is produced by a potential $V_c$, that has a deep minimum on the curve, depends only on the normal coordinates as well as on a parameter $\epsilon$ such that when $\epsilon\rightarrow 0$ the potential goes to infinity outside the submanifold \cite{Costa1981,Maraner1993,Maraner95}. Expanding $V_c$ as a power series
in the $y_i$ about its minimum and assuming that is symmetric in the $y_i$ up to quadratic order \cite{Jaffe2003}, we arrive at\footnote{Asymmetric confining potentials are considered in \cite{Maraner1993,Maraner95}.}
\be
V_c(y^i)=\frac{1}{2\epsilon^4}m\omega y_iy^i+O(y^3).
\ee
Following Refs. \cite{Jaffe2003,Maraner95}, let us rescale the normal coordinates by $y \rightarrow \epsilon y$. Then,  expanding
the Hamiltonian $\mathcal{H}$ up to order $\epsilon^2$, we obtain
\be
\epsilon^2\mathcal{H}=\mathcal{H}^{(0)}+\epsilon\mathcal{H}^{(1)}+\epsilon^2\mathcal{H}^{(2)}+\ldots.
\ee
Since $A_b(s,\epsilon y^i)$ is now dependent on $\epsilon$, it can be Taylor expanded as follows 
\be
A_b(s,\epsilon y^i)=
A_b(s,0)+\epsilon y^j\left.\frac{\partial A_b(s,y^i)}{\partial y^j}\right|_{y=0}+\ldots .
\ee
Thus, the zero, first\footnote{The notations $\bar{A}_b=A_b(s,0)$ and
  $\frac{\partial^n \bar{A}_b}{\partial
    (y^j)^n}=\left.\frac{\partial^n A_b(s,y^i)}{\partial
      (y^j)^n}\right|_{y=0}$ 
have been introduced.} and second order terms are respectively
\be
\mathcal{H}^{(0)}&=&-\frac{\hbar^2}{2m}\partial_i\partial^i+\frac{1}{2}m\omega y_iy^i,\\\label{FirstCouplingS}
\mathcal{H}^{(1)}&=&\frac{ie\hbar}{m} \bar{A}_{i}\partial^{i},
\ee
\be\nn
\mathcal{H}^{(2)}\chi&=&-\frac{\hbar^2}{2m}\hat{\partial_s}\hat{\partial^{s}}\chi
+V_G\chi
+\frac{ie\hbar}{m} \bar{A}_{s}\partial_{s}\chi
+\frac{ie\hbar}{2m}(\partial_{s}\bar{A}_{s})\chi +\frac{ie\hbar}{m} \bar{A}_{s}y^{k}\beta_{k}^{\ j}\partial_{j}\chi\\\label{SecondOrderS}
&&+\frac{e^2}{2m}(\bar{A}_i \bar{A}^i\chi+\bar{A}_{s}\bar{A}^{s})\chi
+\frac{ie\hbar}{2m}\frac{\partial  \bar{A}_i}{\partial y_i}\chi+ \frac{ie\hbar}{m}\frac{\partial  \bar{A}_i}{\partial y^j} y_j\partial^{i}\chi,
\ee
where $\hat{\partial}_{\mu}\equiv\partial_{\mu}+\frac{1}{2\hbar}i\beta_{\mu}^{ij}L_{ij}$, with
$L_{ij}=i\hbar(y_{j}\partial_{i}-y_{i}\partial_{j})$ the angular momentum operators in the space normal to $\mathcal{C}$ and where $V_G$ is a \textit{quantum potential} given by
\be
V_G=-{\frac{\hbar^2}{8m}}\alpha_{k}\alpha^k.
\ee
To obtain an effective dynamics on the submanifold, it is necessary to ``freeze''  the normal degrees
of freedom \cite{Maraner95,Jaffe2003}. However, the first-order term in Eq. (\ref{FirstCouplingS})
is coupling the dynamics of the tangent and normal degrees of freedom through the normal components of the
electromagnetic potential evaluated at $y^j$ equal to zero ($\bar{A}_i=A_i(s,0)$). When the codimension is equal to one it is always possible to choose a gauge in such a way that {$A_{i=n}(s,y^i)=0$}, analogously to how it was done in Ref \cite{Ferrari2008}. In the general case, we see that it is only necessary that $A_i(s,0)=0$ in order to decouple the dynamics. One can always use a gauge transformation $A'_b=A_b+\partial_b \gamma$ so that $\bar{A}_i=0$, namely,
\be
\gamma(s,y^j)=-A_i(s,0)y^i.
\ee
We remark that when the coupling in (\ref{FirstCouplingS}) is $A_i$, rather than $\bar{A}_i$, this is only possible for codimension one.
In general, there is no gauge transformation that eliminates the term $T_{kj}\equiv\left.\frac{\partial A_k(s,y^i)}{\partial y^j}\right|_{y=0}$ unless the codimension is equal to one. However, using the gauge transformation
\be
\gamma'(s,y^j)=-y^ky^j\frac{T_{kj}}{2},
\ee
yields
\be\nn
A'_i&=&A_i+\partial_i \gamma'=y^jT_{ij}-y^j\frac{T_{ij}}{2}-y^j\frac{T_{ji}}{2}+O(y^2)\\
&=&y^j\frac{T_{ij}}{2}-y^j\frac{T_{ji}}{2}+O(y^2)=-\frac{y^j}{2}\bar{F_{ij}},
\ee
\textit{i.e.} the gauge transformation eliminates the symmetric part of $T_{ij}$, leaving only the antisymmetric part which is proportional to electromagnetic tensor evaluated in $y^j$ equal to zero $(\bar{F}_{ij}=\left.F_{ij}/2\right|_{y=0})$. Thus, without loss of generality,
in what follows we will set $\bar{A}_i=0$ and assume consistently that $T_{ij}$ is purely antisymmetric. With these conditions Eq. (\ref{SecondOrderS}) can be written as
\be\label{H2SANT}
\mathcal{H}^{(2)}\chi&=&\bigg[-\frac{\hbar^2}{2m}\left(\partial_{s}+\frac{1}{2\hbar}i\beta^{ij}L_{ij}-\frac{ie}{\hbar} \bar{A}_{s}\right)^2
-e\bar{A}_{0}-{\frac{\hbar^2}{8m}}\alpha_{k}\alpha^k-\frac{e}{2m}\bar{F}^{ij}L_{ij}
\bigg]\chi.
\ee
Let us now consider the separated solution\footnote{Recently \cite{Taira2010a,Taira2010b,Taira2010c},
it has been considered a solution of the form $\chi(t,s,y^i)=\varphi(s,t)\sum_{\beta}\psi_{\beta}(y^i,t)$.
However, this is not the most general case and consequently ignores that wave function in the
low-dimensional space will be a multiplet.}
\be\label{SolFrozenS}
\chi(t,s,y^i)=\sum_{\beta}\varphi_{\beta}(s,t)\psi_{\beta}(y^i,t),
\ee
where the index $\beta$ runs over the degeneracy that exists in the spectrum of the normal degrees
of freedom \cite{Jaffe2003} (this degeneracy is due to the invariance of $\mathcal{H}^{(0)}$ under
rotations of the coordinates $y^1$, $y^2$).  To zeroth-order all the functions $\psi_{\beta}(y^i,t)$ satisfy
\be
\mathcal{H}^{(0)}\psi_{\beta}(y^i,t)=\mathcal{E}\psi_{\beta}(y^i,t).
\ee
Strictly, we should write the solution (\ref{SolFrozenS}) as
\be\label{SolFrozenS1}
\chi^{\mathcal{E}}(t,s,y^i)=\sum_{\beta}\varphi^{\mathcal{E}}_{\beta}(s,t)\psi^{\mathcal{E}}_{\beta}(y^i,t),
\ee
to emphasize that this corresponds to a single energy $\mathcal{E}$. 
{However, one can show, for symmetric confining potentials, that $\mathcal{H}^{(2)}$ can not produce transitions between the energy levels of the normal degrees of freedom} \footnote{{Since that 
$[L_{kl},\mathcal{H}^{(0)}]=0$, one have that $\{\psi^{\mathcal{E}}_{1},\cdots,\psi^{\mathcal{E}}_{\beta}\}$ forms a subspace that is invariant under the action of the normal angular momentum, \textit{i.e.} $L_{kl}\psi^{\mathcal{E}}_{\beta}(y^i,t)=\sum_{\lambda}C_{\beta\lambda}\psi^{\mathcal{E}}_{\lambda}(y^i,t)$.}}. 
Then the system is block diagonal, with each block built with the subspace spanned by $\psi_{\alpha}$. Note that the dimension of each block, $d_e$, depends on the degeneracy that exists in the spectrum of the normal degrees of freedom. 
{Since the ground state of the $\mathcal{H}^{(0)}$ has zero angular momentum, it is necessary prepared the system in an excited state of $\mathcal{H}^{(0)}$ if one want to see associated effects to the angular momentum of the normal degrees of freedom. Without loss of generality, we will assume in what follows that the state is prepared in an eigenstate of $\mathcal{H}^{(0)}$ and so omit the index $\mathcal{E}$. Inserting (\ref{SolFrozenS1}) into Eq. (\ref{H2SANT}), multiplying  by $\psi^{*}_{\alpha}(y^i)$ and integrating in $\int dy^1dy^2$, one can be write a Schrödinger equation for the tangent degrees of freedom as follows}\footnote{Without loss of generality it was used that $\int dy^1dy^2\psi_{\beta}^{*}(y^i)\psi_{\alpha}(y^i)=\delta_{\beta \alpha}$.}
\be\nn
i\hbar\frac{\partial \vec{\varphi}(s,t)}{\partial t}&=&\bigg[-\frac{\hbar^2}{2m}\left(\partial_{s}+\frac{1}{2\hbar}i\beta^{ij}
\mathcal{L}_{ij}-\frac{ie}{\hbar} \bar{A}_{s}\right)^2
-e\bar{A}_{0}\bigg.\\\label{EfecConW}
&&\bigg.-{\frac{\hbar^2}{8m}}\alpha_{k}\alpha^k-\frac{e}{2m}\bar{F}^{ij}
\mathcal{L}_{ij}+{\frac{1}{8m}}\beta^{ij}\beta^{kl}\left({\mathcal{L}^2_{ij,kl}}-\mathcal{L}_{ij}\mathcal{L}_{kl}\right)
\bigg]\vec{\varphi}(s,t),
\ee
where {$\vec{\varphi}(s,t)$} is a multiplet $\vec{\varphi}=(\varphi_1,\ldots,\varphi_{d_e})$
and the matrices $\mathcal{L}_{ij}$ and $\mathcal{L}_{ij,jk}$ are defined as
\be
(\mathcal{L}_{ij})_{\alpha\beta}&\equiv&\int dy^1dy^2\psi^{*}_{\alpha}(y)L_{ij}\psi_{\beta}(y),\\
{(\mathcal{L}^2_{ij,kl})_{\alpha\beta}}&\equiv&\int dy^1dy^2\psi^{*}_{\alpha}(y)L_{ij}L_{kl}\psi_{\beta}(y).
\ee
The last term in Eq. (\ref{EfecConW}) has been taken into account recently in the study of electrons in a 
twisted quantum ring \cite{Taira2010a,Taira2010a}.
However, it has been previously pointed out in the literature \cite{Jaffe2003} that since the confining 
potential $V_c$ is invariant under rotations of the coordinates $y^i$, all of the $\mathcal{L}_{ij}$s 
commute with $\mathcal{H}^{(0)}$ and $\{\psi_{\beta}\}$ forms a complete set
of states for the subspace associated with a fixed level of energy $\mathcal{E}$. Consequently,
the last term in Eq. (\ref{EfecConW}) vanishes identically\footnote{This term can be different from zero if $V_c$ is asymmetric \cite{Maraner1993,Maraner95}.}.
\par
The term proportional to $\bar{F}^{ij}\mathcal{L}_{ij}$ can be written as
\be
-\frac{e}{2m}\bar{F}^{ij}\mathcal{L}_{ij}=-\frac{e}{m}\bar{F}^{12}\mathcal{L}_{12}=\frac{2}{\hbar}B_s\mu \, \mathcal{S}
\ee
with $\mu=-e\hbar/(2m)$, $B_s=\bar F^{12}$ the magnetic field tangent to $\mathcal{C}$ and defining  $\mathcal{S}=\mathcal{L}_{12}$ as the only independent component of the matrix $\mathcal{L}_{ij}$. Taking into account
that $\beta_{ij}$ is antisymmetric, one can define the \textit{torsion} $\tau\equiv\hat{\n}_2\cdot\partial_x \hat{\n}_1=\beta_{21}$ as the only independent component of $\beta_{ij}$ and one can be rewritten (\ref{EfecConW}) as
\be\label{SchOneBS}
i\hbar\frac{\partial \vec{\varphi}(s,t)}{\partial t}&=&
\bigg[-\frac{\hbar^2}{2m}\left(\partial_{s}-\frac{i\tau \mathcal{S}}{\hbar}-\frac{ie}{\hbar} \bar{A}_{s}\right)^2
-e\bar{A}_{0}-{\frac{\hbar^2}{8m}}\alpha_{k}\alpha^k+\frac{2}{\hbar}B_s \mu\mathcal{S}
\bigg]\vec{\varphi}(s,t).
\ee
The last term in (\ref{SchOneBS}) is a new quantum potential, which is independent of the
curvature and torsion of the curve and, therefore, is present even in the line. This extends
previous effects found in the literature due to dimensional reduction, in particular, the results of
Refs. \cite{Ferrari2008,Pershin2005a} where only were studied cases when the dimension is reduced
by one. In our case, an ``inner'' observer perceives the orbital angular momentum of the normal coordinates
as an ``intrinsic angular momentum''. Thus, one can interpret the new quantum potential as an induced
Zeeman interaction where the \textit{induced magnetic moment} is\footnote{Note that there is
no orbital angular momentum in a one-dimensional system.}
\be
\mathcal{\mu}_s=-\frac{2}{\hbar}\mu\mathcal{S},
\ee
with a induced Landé $g$-factor $g_s=2$. In a pure one-dimensional system the electromagnetic potential is
$A_\mu=(A_0,A_s)$ and there is no magnetic field (note that in Eq. (\ref{SchOneBS}) the magnetic field
$B_s=\partial_{y^1}A_{y^2}-\partial_{y^2}A_{y^1}$ is independent of the component $A_s$).
Comparing Eq. (\ref{SchOneBS}) with the one-dimensional Schrödinger equation in the presence of electromagnetic fields
\be
i\hbar\frac{\partial \varphi(s,t)}{\partial t}=
\bigg[-\frac{\hbar^2}{2m}\left(\partial_{s}-\frac{ie}{\hbar} \bar{A}_{s}\right)^2
-e\bar{A}_{0}\bigg]\varphi(s,t),
\ee
we see that the dimensional reduction produces the well known geometric potential, a geometry-induced
Yang-Mills field and, in the presence of magnetic fields, the induced Zeeman coupling. All
these effects must be considered when working on systems such as quantum rings or quantum wires.
\par
Let us consider briefly what happens to non-relativistic particles with spin. In this case, the equation is the Pauli equation
that can be written as
\be\label{PauElcCov}
i\hbar \D_{0}\psi(t,s,y^i)=\left[-\frac{\hbar^2}{2m|G|^{1/2}}D_a
|G|^{1/2}G^{ab}D_b+V_c(y^i)-\frac{e}{m}\mathbf{B}\cdot \mathbf{S}\right]\psi(t,s,y^i),
\ee
where $\psi$ has two components and $\mathbf{S}=\hbar\mathbf{\sigma}/2$. It is clear that to analyze the Pauli equation (\ref{PauElcCov})
only need to consider the last term. Proceeding as before, we arrive at
\be\nn
i\hbar\frac{\partial \vec{\varphi}(s,t)}{\partial t}&=&
\bigg[-\frac{\hbar^2}{2m}\left(\partial_{s}-\frac{i\tau \mathcal{S}}{\hbar}-\frac{ie}{\hbar} \bar{A}_{s}\right)^2
-e\bar{A}_{0}-{\frac{\hbar^2}{8m}}\alpha_{k}\alpha^k-\frac{\hbar e}{m}(B_s \mathcal{S}
+\bar{\mathbf{B}}\cdot \mathbf{S})\bigg]\vec{\varphi}(s,t),
\ee
where $\bar{\mathbf{B}}=\mathbf{B}(s,0)$ and now $\vec{\varphi}$ has $2\times d_e$ components, \textit{i.e.}
$\vec{\varphi}$ has $d_e$ components each one with spin.
\subsection{Example 1: Untwisted quantum ring}
Let us show a simple example, a ring crossed by an infinite wire carrying current $I$. The magnetic field lines
are tangential to the ring with an intensity of $B_s=\frac{\mu_0I}{2\pi R}$, with $R$ the
radius of the ring. The effective Schrödinger equation
(\ref{SchOneBS}) yields the following equation
\be
E\vec{\varphi}^{\mathcal{E}}&=&\bigg[-\mathbf{1}\frac{\hbar^2}{2m}\partial_{s}^2
-\mathbf{1}{\frac{\hbar^2}{8m}}\alpha_{k}\alpha^k+\frac{2}{\hbar}B_s \mu\mathcal{S}
\bigg]\vec{\varphi}^{\mathcal{E}},
\ee
where we put the subscript $\mathcal{E}$ again to remember that this state depends on the energy levels
of $\mathcal{H}^{(0)}$ and with $\mathbf{1}$ the $d_e\times d_e$ identity matrix. 
{If the system is prepared in} the ground state of
$\mathcal{H}^{(0)}$, the angular momentum in the normal coordinates ($\mathcal{S}$) is zero. Thus,
\be
E_l=\frac{\hbar^{2}}{2mR^2}l^2-\frac{\hbar^{2}}{8mR^2}, \ \ \ \ l=0,\pm 1,\pm 2, \cdots,
\ee
note that except for the ground state all levels has a 2-fold degeneracy. 
{Now,  for a system prepared in the first excited state of
$\mathcal{H}^{(0)}$, one have that $\vec{\varphi}=(\varphi_1,\varphi_2)$ 
due to that $\mathcal{H}^{(0)}$ is doubly degenerate.} 
One calculate that in this level $\mathcal{S}=diag(\hbar,-\hbar)$, so
\be
E_l=\frac{\hbar^{2}}{2mR^2}l^2-\frac{\hbar^{2}}{8mR^2}\pm 2B_s\mu, \ \ \ \ l=0,\pm 1,\pm 2, \cdots,
\ee
with $+$ for $\varphi_1$ and $-$ for $\varphi_2$. When $B_s\neq0$ three is a split in the spectrum, otherwise,
the ground state is 2-fold degenerate and all other states are 4-fold degenerate.
{
\subsection{Example 2: Helix} Let us consider particles confined in a helix parameterized as $\vec{r}=(R\cos s/\sqrt{R^2+c^2},R\sin s/\sqrt{R^2+c^2},c)$, where $s$ is the arc-length. As in the last example, the helix is crossed by an infinite wire carrying current $I$. One can show that the time-independent Schrödinger equation is given by 
\be\nn
E\vec{\varphi}^{\mathcal{E}}&=&
\bigg[-\frac{\hbar^2}{2m}\left(\partial_s-\frac{ic \mathcal{S}}{(R^2+c^2)\hbar}+\frac{iec\mu_0 I Ln(R/R_0)}{2 \pi\hbar\sqrt{R^2+c^2}} \right)^2\\
&&-\frac{\hbar^2R^2}{8m(R^2+c^2)^2}+\frac{2}{\hbar}\frac{\mu_0I}{2\pi\sqrt{R^2+c^2}} \mu\mathcal{S}
\bigg]\vec{\varphi}^{\mathcal{E}}.
\ee
If the system is in the first excited state
of $\mathcal{H}^{(0)}$, the energy the spectrum is continuous and is given by  
\be
E_p=\frac{1}{2m}\left(p_s\pm\frac{c\hbar}{(R^2+c^2)}-\frac{ec\mu_0 I Ln(R/R_0)}{2 \pi\sqrt{R^2+c^2}} \right)^2-\frac{\hbar^2R^2}{8m(R^2+c^2)^2}\pm\frac{\mu_0I}{\pi\sqrt{R^2+c^2}} \mu
\ee
with $+$ for $\varphi_1$ and $-$ for $\varphi_2$. 
The above example clearly illustrates that, in general, is important to consider the effect of torsion, curvature and the induced Zeeman coupling in the dynamics of particles in one-dimensional systems. }

\section{Conclusion}\label{Section5KG}
We have employed the thin-layer method to determine the effective dynamics of
a spinless and $1/2$-spin particles constrained on a space curve and
in the presence of electromagnetic fields. We have found that there are no
coupling between the dynamics of the normal and tangent degrees of freedom
through the normal components of the electromagnetic potential. We report a
new quantum potential, which can be interpreted as an induced Zeeman coupling.
This new effect needs to be considered when working on systems such as
quantum rings or quantum wires in the presence of magnetic fields. {Therefore,}
this paper extends previous results \cite{Ferrari2008,Pershin2005a}
in which spinless charged particles were studied when the dimension is reduced
by one. The effect of reducing the dimension by two allows to identify an 
``induced intrinsic angular momentum'' and an induced magnetic moment.

\acknowledgments
J. A. Sánchez would like to thank CAPES, Brazil,  for financial
support. F. T. Brandt would like to thank CNPq for a grant.

\end{document}